\documentclass[fleqn,10pt]{wlscirep}
\usepackage[utf8]{inputenc}
\usepackage[T1]{fontenc}
\usepackage{amsmath}

\title{Applying the Fokker--Planck equation to grating-based x-ray phase and dark-field imaging}

\author[1,2,*]{Kaye S. Morgan}
\author[1]{David M. Paganin}
\affil[1]{School of Physics and Astronomy, Monash University, Clayton, Victoria, 3800, Australia}
\affil[2]{Chair of Biomedical Physics, Department of Physics, Munich School of Bioengineering, and Institute of Advanced Study, Technische Universit\"{a}t M\"{u}nchen, 85748, Garching, Germany}

\affil[*]{Kaye.Morgan@monash.edu}


\begin{abstract}
X-ray imaging has conventionally relied upon attenuation to provide contrast.  In recent years, two complementary modalities have been added; a) phase contrast, which can capture low-density samples that are difficult to see using attenuation, and b) dark-field x-ray imaging, which reveals the presence of sub-pixel sample structures.   These three modalities can be accessed using a crystal analyser, a grating interferometer or by looking at a directly-resolved grid, grating or speckle pattern.  Grating and grid-based methods extract a differential phase signal by measuring how far a feature in the illumination has been shifted transversely due to the presence of a sample.  The dark-field signal is extracted by measuring how the visibility of the structured illumination is decreased, typically due to the presence of sub-pixel structures in a sample.  The strength of the dark-field signal may depend on the grating period, the pixel size and the set-up distances, and additional dark-field signal contributions may be seen as a result of strong phase effects or other factors.  In this paper we show that the finite-difference form of the Fokker--Planck equation can be applied to describe the drift (phase signal) and diffusion (dark-field signal) of the periodic or structured illumination used in phase contrast x-ray imaging with gratings, in order to better understand any cross-talk between attenuation, phase and dark-field x-ray signals. In future work, this mathematical description could be used as a basis for new approaches to the inverse problem of recovering both phase and dark-field information.
\end{abstract}
\begin{document}

\flushbottom
\maketitle

\thispagestyle{empty}

\section{Introduction}

X-ray imaging has been widely adopted in medicine and security, providing non-invasive visualisation of internal structure at high spatial and temporal resolution.  The conventional mechanism for x-ray contrast is attenuation, where dense and/or thick objects reduce the intensity of the illuminating x-ray wavefield.  More recently, the capabilities of x-ray imaging have been extended in research laboratories and synchrotrons, capturing weakly-attenuating features like soft biological tissue \cite{momose1996} through measurements of the wavefield phase, and revealing sub-pixel structures like dentinal tubules \cite{jud2016dentinal} via a `dark-field' signal that originates from unresolved scattering.  Because an x-ray detector is typically only sensitive to wavefield intensity,  specialised optical set-ups are required to convert variations in wavefield phase and dark-field into measurable variations in intensity. A mathematical model of the behaviour can then be used to retrieve the phase and dark-field information, typically from several exposures taken with different optical settings.  There are a number of set-ups that have demonstrated the ability to do this, beginning with crystal interferometry \cite{BonseAndHart} in the 1960s.  The advent of highly coherent synchrotron x-ray sources several decades later led to propagation-based phase contrast x-ray imaging, where the coherent wavefield self-interferes during propagation to reveal phase information in addition to attenuation effects \cite{Cloetens, Snigirev}. More recently a number of techniques have been developed that utilise high-frequency gratings to measure attenuation, phase and dark-field information.  These techniques include grating interferometry \cite{Weitkamp, David, Pfeiffer2008df, Momose, miao2016universal}, edge-illumination \cite{olivo2007coded} and single-grating imaging \cite{Bennett2010, morgan2011quantitative}. A generalisation of single-grating imaging is speckle-tracking \cite{Morgan2012, berujon2012}, where the grating is substituted with a object, typically a sheet of sandpaper, that creates a reference pattern that need not be periodic, and is typically random.   

\begin{figure}
\includegraphics[width=500pt]{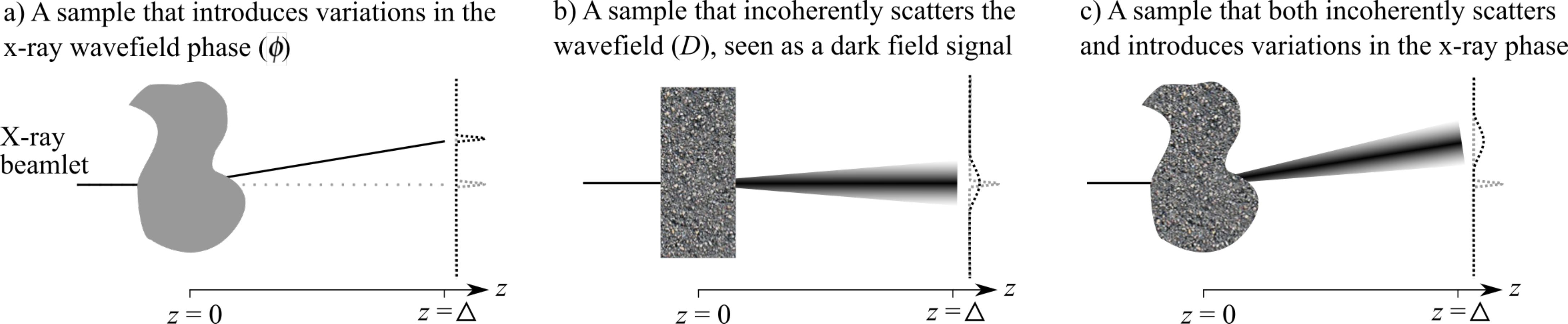}
\caption{A photon or x-ray beamlet may encounter: a) a phase gradient in the sample, transversely shifting the position of the detected light at a downstream detector, b) small angle scattering from the sample, spreading out the beamlet, reducing the visibility, and broadening the probability distribution that describes where the photon will land at a downstream detector, or c) both effects.}
\label{fig:Samples}
\end{figure}

This manuscript focuses on the grating and speckle-based phase contrast x-ray imaging set-ups. These set-ups can extract an attenuation image, a differential phase contrast image (or, in some cases, two complementary directions of differential phase contrast), and a dark-field image (or directional dark-field images in some cases, when applied to a sample with aligned elongated sub-pixel features). These image modalities can be understood by considering a pencil beam, or beamlet, passing through the sample. This beamlet will decrease in total intensity if the sample is attenuating.  Variations in the thickness or refractive index of the sample can alter the phase of the x-ray wave, and result in a change in the direction of beamlet propagation, as seen in Fig.~\ref{fig:Samples}a. Therefore, the phase signal can be extracted by looking at some kind of transverse shift in the intensity profile at a distance downstream (e.g. $z=\Delta$, where $z$ is the optic axis).  If the beamlet is symmetrically increased in width, as a result of sub-resolution sample structures (Fig.~\ref{fig:Samples}b) \cite{khelashvili2005}, this is described as a scatter, visibility or dark-field signal, measured by looking at a reduction in the visibility of the illumination \cite{Endrizzi2018}.  Visibility can be defined using Michelson's definition $V=(I_{max}-I_{min})/(I_{max}+I_{min})$, where $I_{max}$ is the local maximum intensity and $I_{min}$ is the local minimum intensity \cite{Michelson95}, so that high visibility indicates that $I_{max}$ is much larger than $I_{min}$. A strong dark-field signal will be seen as a significant decrease in illumination visibility. The dark-field signal measured by a grating-based system originates largely from small angle x-ray scattering (SAXS) or incoherent scattering that is below the resolution of the imaging system, but can also result from edges \cite{yashiro2015} or focusing \cite{wolf2015lens}, beam hardening \cite{chabior2011beam, zdora2015simulations} and other effects.  In most imaging applications, the sample will introduce both a transverse shift and broadening of the beamlet at a distance $z=\Delta$, as seen in Fig.~\ref{fig:Samples}c.  The complementary transverse directions (e.g. $x$, $y$) of phase contrast and directional dark-field would be measured by looking at transverse shifts and broadening in the direction shown in the diagram and the direction perpendicular to the page.

The intensity profiles shown on the right of each panel in Fig.~\ref{fig:Samples} are reminiscent of a probability density function that experiences changes in mean (Fig.~\ref{fig:Samples}a) and standard deviation (Fig.~\ref{fig:Samples}b).  The sum of x-ray photons within each pixel, measured as intensity, can be modelled as a probability density function since the integrated intensity from sample exit plane to downstream plane is constant for a beam propagating in vacuum. This paper therefore studies the Fokker--Planck equation, an equation typically used to describe the time-evolution of a probability density function, and applies it to describe the propagation-evolution of a beamlet in x-ray phase/dark-field imaging \cite{Risken1989}. In a companion paper\cite{Paganin2019}, we show how the Fokker--Planck equation is an extension to the transport of intensity equation \cite{Teague1983} used in coherent x-ray optics and also provide a derivation from first principles. Here, we begin with a short probability-based introduction to the Fokker--Planck equation (Sec. \ref{sec:Probability}), then show that it can be applied to better understand how the x-ray dark-field (Sec. \ref{sec:DF}) and differential phase (Sec. \ref{sec:Phase}) signals can interact, looking at additional effects seen near edges and as a result of focusing.  These two sections (\ref{sec:DF} and \ref{sec:Phase}) look specifically at imaging using patterned illumination that is directly resolved (e.g. single-grid imaging with an attenuating \cite{Bennett2010, Morgan2010a} or phase-shifting grating \cite{Morgan2013, rizzi2013}, or speckle-tracking \cite{Morgan2012, berujon2012, zdora2018}).  Given that attenuation is a simple scaling of the intensity signal, hence a multiplicative factor at the front of the Fokker--Planck equation, we leave out this effect for simplicity. 
The final section, \ref{sec:InterferometryEdgeIllumination}, demonstrates that the results and approach described in the earlier sections can be extended to the cases where the illumination is analysed via a second grating or mask (e.g. grating interferometry or edge-illumination imaging).

\section{The X-ray Fokker--Planck Equation}
\label{sec:Probability}

\begin{figure}
\includegraphics[width=500pt]{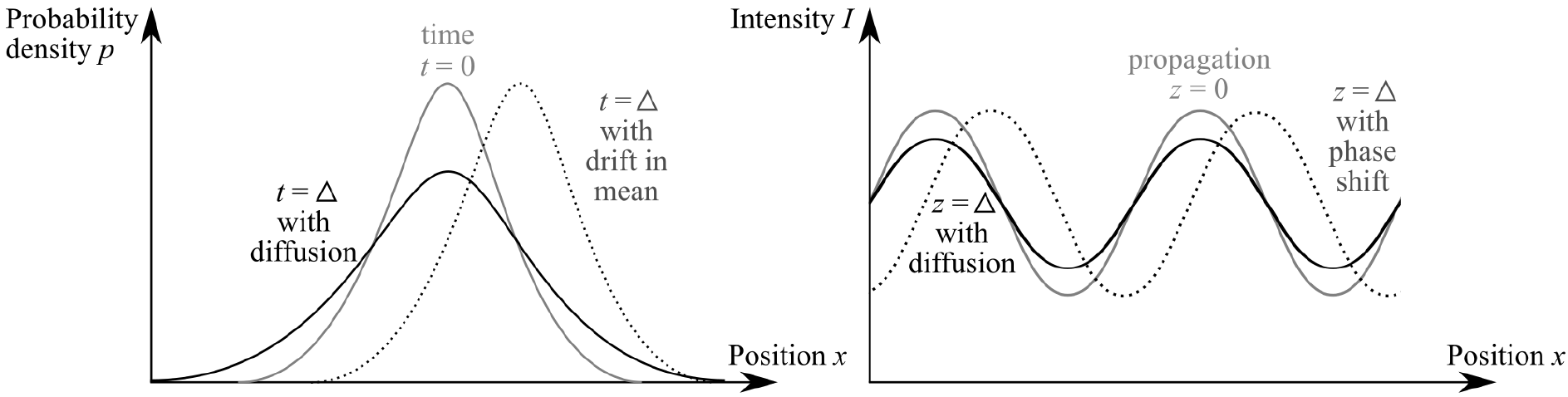}
\caption{a) The Kramers--Moyal equation and its truncated equivalent, the Fokker--Planck equation, describe the diffusion and shift in a probability function. b) This same equation can be applied to study the x-ray dark-field and phase shift measured using periodic x-ray illumination.}
\label{fig:FokkerPlanck}
\end{figure}

The Fokker--Planck equation, shown below, describes the evolution of a probability density function $p$ with time $t$. Changes can include a time-dependent drift in both mean ($D_1$, a transport-type energy flow) and/or standard deviation ($D_2$, a diffusive-type energy flow) \cite{MandelWolf},
\begin{equation}
\label{eq:FokkerPlanck}
\frac{\partial}{\partial t} p(x,t) = - \frac{\partial}{\partial x} (D_1(x,t) p(x,t)) + \frac{\partial^2}{\partial x^2} (D_2(x,t) p(x,t)),
\end{equation}

\noindent as shown in Fig.~\ref{fig:FokkerPlanck} a after a given time interval $t=\Delta$. 

Since the path taken by an ensemble of photons will follow a probability density function and be seen as such when those locations are registered as intensity on the detector, we can use the Fokker--Planck equation to model changes in the measured x-ray intensity. The Fokker--Planck equation can also be seen as the transport of intensity equation (TIE) with the addition of a diffusive term \cite{Paganin2019}. In the grating-based phase contrast x-ray imaging methods mentioned above, a grating will typically split the beam into an array of `beamlets', each of which will pass through the sample and propagate through free space to a detector.  We can begin by looking at the one-dimensional intensity distribution measured downstream of a sample that is illuminated by an array of beamlets, with a transverse shift of each beamlet seen at $z=\Delta$ when the beamlet passes through a phase-shifting sample (Fig.~\ref{fig:Samples} a), and a reduction in visibility of each beamlet seen at $z=\Delta$ when the beamlet scatters at small angles from sub-pixel features in the sample (Fig.~\ref{fig:Samples} b).   

To move from the time-evolution of a probability function (Fig.~\ref{fig:FokkerPlanck}a, Eqn. (\ref{eq:FokkerPlanck})) to the case of spatial drift and diffusion of an array of x-ray beamlets in the presence of a sample (Fig.~\ref{fig:FokkerPlanck}b,  Eqn.~(\ref{eq:FokkerPlanckXray}), where $I(x)$ is a sinusoid), we should replace:

\begin{itemize}
\item Time $t$ with propagation distance $z$
\item Probability density function $p$ with x-ray intensity $I$
\newline(In the case shown in Fig.~\ref{fig:FokkerPlanck}, the centred probability density function $p$ is replaced by an array of x-ray intensity beamlets $I$, described by a sinusoid)
\item Drift velocity $D_1$ with the change in angle of x-ray wavefield propagation, determined by the sample-induced phase shift $\frac{1}{k} \frac{\partial \phi}{\partial x}$
\item Diffusion coefficient $D_2$ with a value $D(z)$ that characterises the length scale of unresolved x-ray scattering at distance $z$.
\end{itemize}

This gives the `X-ray Fokker--Planck equation' \cite{Paganin2019},

\begin{equation}
\label{eq:FokkerPlanckXray}
\frac{\partial}{\partial z} I (x) = -\frac{1}{k} \frac{\partial}{\partial x} \left(I \frac{\partial \phi (x)}{\partial x}\right) + \frac{\partial^2}{\partial x^2} (D(x,z) I(x)).
\end{equation}

Note that the conventional Fokker--Planck model will describe a diffusion that increases the spread of light in a non-linear manner with propagation. This is not consistent with the linear increase in spread with propagation distance seen when a scattered wave-field propagates through free space.   For this reason, $D(x,z)$ is set as a function of $z$, so that the diffusion width can increase linearly with propagation distance $z$ (see Paganin \& Morgan for further justification\cite{Paganin2019}). 

In order to fully describe directional, three-dimensional or more complex behaviour, we would need an equation that includes more spatial dimensions and terms.  This is a good time to note that the Fokker--Planck equation is the truncated form of the more general Kramers--Moyal equation \cite{Risken1989,MandelWolf}. This more general equation includes third-order and higher-order terms, and describes changes in $n$ spatial dimensions, with \noindent ${\bf{x}}=(x_1,x_2,x_3...)$ as Cartesian coordinates, as below,

\begin{equation}
\label{eq:KramersMoyal}
\frac{\partial}{\partial t} p(\textbf{x},t) = \sum_{r=1}^{\infty} \frac{(-1)^r}{r!}\frac{\partial^r}{\partial x^r} \left[D_r(\textbf{x},t)p(x,t)\right]
=-\frac{\partial}{\partial x_i}\left[D_i(\textbf{x},t)p(\textbf{x},t)\right]+\frac{1}{2!}\frac{\partial^2}{\partial x_i \partial x_j}\left[D_{ij}(\textbf{x},t)p(\textbf{x},t)\right]-\cdots
\end{equation}

For the case of two dimensions and truncation immediately before the `$-...$' shown above, we have the anisotropic-diffusion Fokker-Planck equation, which well-describes a two-dimensional x-ray beamlet landing on a two-dimensional detector,

\begin{equation}
\label{eq:KramersMoyalXray}
\frac{\partial}{\partial z} I(\textbf{x},z) = - \frac{\partial}{\partial x} \left(\frac{1}{k} \frac{\partial \phi}{\partial x} I(\textbf{x},z)\right) -\frac{\partial}{\partial y} \left(\frac{1}{k} \frac{\partial \phi}{\partial y} I(\textbf{x},z)\right) + \frac{\partial^2}{\partial x^2} (D_x I(\textbf{x},z)) + \frac{\partial^2}{\partial y^2} (D_y I(\textbf{x},z)) + \frac{\partial^2}{\partial x \partial y} (D_{x,y} I(\textbf{x},z))
\end{equation}

\noindent (absorbing any multiplying fractions into $D_{x}, D_{y}$ and $D_{xy}$). This kind of diffusion in multiple directions is ideal for modelling a directional and/or irregularly shaped x-ray dark-field signal\cite{jensen2010directional}, and is an avenue for future work. The parameters $D_{x}, D_{y}$ and $D_{xy}$ describe a diffusion profile that has the shape of an ellipse, and hence requires three parameters to describe it fully.  These three parameters may be mapped one-to-one to the semi-major axis, the semi-minor axis, and the angular orientation of the elliptical diffusion profile.  Note also, that if the infinitely-many terms in the Kramers--Moyal equation are truncated to no higher than second-order spatial derivatives, the result is the Fokker--Planck Equation.  As shown by Pawula, one should either truncate to second-order or not at all, since `if one assumes that terms above a given order are zero in the Kramers--Moyal expansion, this assumption implies that {\em all} terms above second order are zero'\cite{Pawula1967}.

\section{Application of the Fokker--Planck model to describe phase and dark-field effects}
\label{sec:FPApplication}

To model x-ray grating-based phase and dark-field imaging, the key method considered in the present paper, we can set the intensity that illuminates the sample to a sinusoid. This can describe the intensity that would be seen downstream of a grating or grid, and can be specified for period $p$, amplitude $a$ and mean intensity $b$, 

\begin{equation}
\label{eq:Illumination}
I=a\sin\left(\frac{x}{p}\right)+b.
\end{equation}

Note that we have omitted the factor of $2\pi$ inside the sinusoid that would be required to describe a period of exactly $p$.  This choice was made to keep the equations as simple as possible without neglecting any effects. The general expression given in Eqn. \ref{eq:Illumination} means the illumination without any sample in the beam would have a mean intensity of $b$ and visibility of $a/b$ (e.g. the grey curve in Fig.~\ref{fig:darkfield} a).  Given that a Fourier series could be used to form any other illumination (e.g. periodic `square' illumination sampled an order of magnitude more finely than the grid period \cite{Morgan2010a} or a speckle pattern generated by a piece of sandpaper \cite{Morgan2012, berujon2012, zdora2018}), the choice of a sinusoidal illumination is generally useful. In the illustrations given in Figs. \ref{fig:darkfield}-\ref{fig:phasecurvature}, we have sketched a sinusoidal illumination that is likely seen at a detector downstream of an absorption or phase grid, effectively a normalised intensity (reaching from 0 to 1) that has been smoothed to a more realistic visibility of 0.5 so that the sinusoid is described by $a\approx\frac{1}{4}$ and $b\approx\frac{1}{2}$. A visibility that is less than 1 is typically seen due to limited coherence, some penumbral blur and/or sampling of relatively fine grids (e.g. 7 pixels per grid period) that are chosen to maximise the resolution of the retrieved images \cite{Bennett2010}.

To examine measurements at a given detector position, we can take the x-ray Fokker--Planck equation (Eqn.~(\ref{eq:FokkerPlanckXray})) with the finite-difference approximation for propagation to $z=\Delta$ (functional dependencies, where obvious, have been dropped for simplicity and clarity),

\begin{equation}
\label{eq:TaylorApprox}
I(x,z=\Delta) \approx I(x,z=0) + \Delta \left(-\frac{1}{k}\frac{\partial}{\partial x}\left(I\frac{\partial \phi}{\partial x}\right)+\frac{\partial^2}{\partial x^2} (D I)\right).
\end{equation}

In the following sections we will consider how these two steps (Eqns.~(\ref{eq:Illumination}) and (\ref{eq:TaylorApprox})) can be applied to the dark-field/diffusion term (Section \ref{sec:DF}) and the phase/drift term (Section \ref{sec:Phase}) to show how the Fokker--Planck equation is relevant to grating-based x-ray imaging and to understand how spatial variations in the phase and dark-field may introduce signal variations that could be `falsely' interpreted as attenuation, phase or dark-field signals.  The Supplementary material includes python code for interactive visualisation of the less-intuitive mathematical expressions from Section \ref{sec:FPApplication}, which can be used to more intuitively understand the effects described herein (also found at https://github.com/KayeMorgan7/XrayFokkerPlanckSim).

\subsection{Dark-field / Scattering effects}
\label{sec:DF}

We look first at an object with scattering properties, but no phase-shifting properties ($\partial\phi/\partial x=0$), which will leave only the second term on the right side of the x-ray Fokker--Planck equation (Eqn.~(\ref{eq:FokkerPlanckXray})).  We can insert the sinusoidal illumination $I(x,z)$ (Eqn.~(\ref{eq:Illumination})) and take the finite-difference approximation for Eqn.~(\ref{eq:TaylorApprox}) to leave,

\begin{equation}
 \label{eq:DF}
I(x,z=\Delta) \approx \left(a \sin{\left(\frac{x}{p}\right)} + b \right) + \Delta \frac{\partial^2}{\partial x^2} \left(D  \left(a \sin{\left(\frac{x}{p}\right)} + b\right)\right).
 \end{equation}

\subsubsection{Uniform scattering}
\label{sec:UniformScat}
If we consider an area where the diffusion properties do not vary quickly with $x$ (so that $D$ can be moved outside of $\frac{\partial^2}{\partial x^2}$), we can easily differentiate the sinusoid twice with respect to $x$, and are left with

\begin{eqnarray}
\label{eq:SimpleDiffusion}
I(x,z=\Delta) = \left(1-\frac{D\Delta}{p^2}\right)a \sin\left(\frac{x}{p}\right) + b.
\end{eqnarray}

\begin{figure}
\centering
\includegraphics[width=500pt]{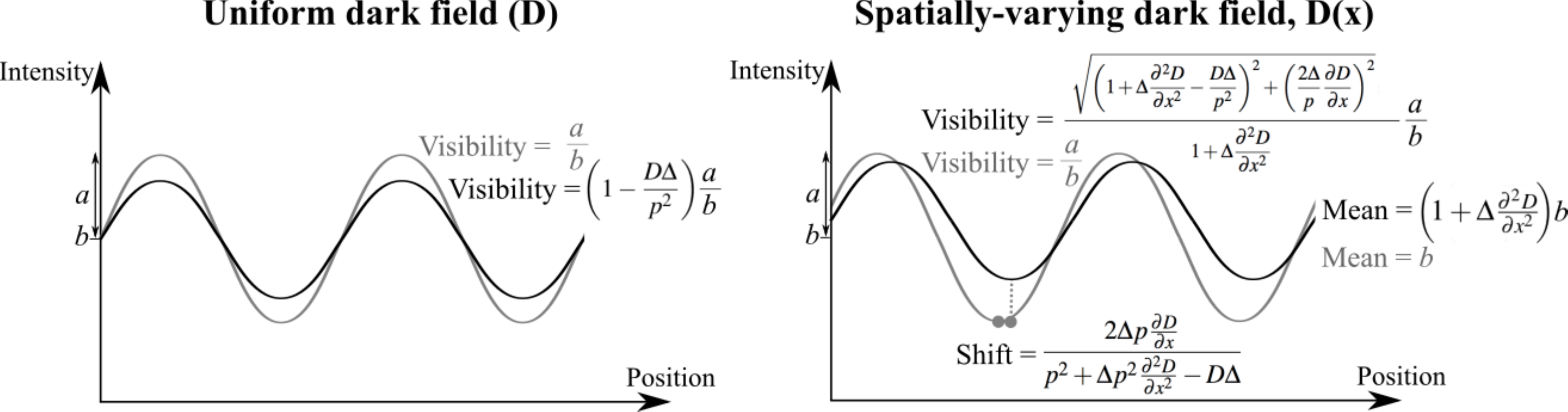}
\caption{a) A diffusion of the illumination is the basis for a dark-field signal, shown here for the simple case where the diffusion $D$ is constant. The reduction in the visibility of the periodic illumination is described by Eqn.~(\ref{eq:SimpleDiffusion}). b) When the diffusion $D$ is not constant, any sudden spatial change in the scattering properties (i.e. $\partial^2D/\partial x^2$ is significant) results in an effect that would be interpreted as variations in dark-field, phase and attenuation, as described in Eqn.~(\ref{eq:DiffEdgeSimp}).}
\label{fig:darkfield}
\end{figure}

Equation~\ref{eq:SimpleDiffusion} describes a decrease in the amplitude of the sinusoid, and hence the visibility of the sinusoid, as seen in Fig.~\ref{fig:darkfield} a), which is indeed the signal which is collected in dark-field imaging.  A more significant effect will be seen for a larger diffusion coefficient $D$, larger propagation distance $\Delta$ or a smaller grid period $p$, as expected. 

Evaluating a direct solution like this can encounter problems if the amplitude of the sinusoidal illumination becomes negative. The problem arises because a linear approximation has been made with propagation distance $\Delta$ (Eqn. (\ref{eq:TaylorApprox})), and the linearisation eventually reaches zero and then becomes negative. Equation (\ref{eq:SimpleDiffusion}) will therefore be valid only when $D \Delta<p^2$.  The case when the intensity sinusoid will reach an amplitude of zero is at $D \Delta/p^2=1$, when the dark-field blurring is over a length scale at the detector $L$ that is equal to the period of the sinusoid $p$. This naturally provides a definition for the `blur width' $L$, $L=\sqrt{D \Delta}$, matching the definition found in the companion paper\cite{Paganin2019}.  Note that if we incorporate $D(z)=D_0 \Delta$, then $L=\sqrt{D_0} \Delta$.

To avoid the limited applicability in any numerical simulations, an alternative solution to the diffusion equation is via a linear integral transform (which can loosely be considered as convolution with a position-dependent point spread function, where the point spread function in our case is a Gaussian where the standard deviation grows with $\sqrt{D\Delta}$)

\begin{equation}
\label{eq:PosDepPSF}
	I(z=\Delta)=\int_{-\infty }^{\infty} I(x',z=0) \frac{e^{-\frac{(x-x')^2}{2 \times D(x')\Delta}}}{\sqrt{2 \pi \times D(x')\Delta}} dx'.
\end{equation}

For sinusoidal illumination, this gives,
\begin{equation}
\label{eq:PosDepPSFUniform}
	I(z=\Delta)=a e^{-\frac{D \Delta }{2p^2}} \sin \left(\frac{x}{p}\right)+b.
\end{equation}

\noindent Note the standard deviation of $\sqrt{D\Delta}$ used in Eqn. \ref{eq:PosDepPSF} could be scaled with a factor of $\sqrt{2}$ to better match the blurring width used in the finite-difference approximation so that Eqns. \ref{eq:SimpleDiffusion} and \ref{eq:PosDepPSFUniform} match at low propagation distance $z=\Delta$, but the approach here is sufficient for the order of magnitude effects in modelling the diffusive behaviour.

\subsubsection{Scattering edge effects}
If the diffusion property $D$ does change quickly with $x$, such that $D$ cannot be moved outside $\frac{\partial^2}{\partial x^2}$ in Eqn.~(\ref{eq:DF}), then we start to see more interesting behaviour.  This case is most likely to correspond to an edge of a strongly-scattering region.  We can expand the brackets from Eqn.~(\ref{eq:DF}) to get

 \begin{equation}
 \label{eq:ScatEdge}
 I(x,z=\Delta) = a\sin\left(\frac{x}{p}\right) + b + \Delta \left[\frac{\partial^2D}{\partial x^2}\left(a\sin\left(\frac{x}{p}\right)+b\right)-\frac{D}{p^2}a\sin\left(\frac{x}{p}\right)+\frac{2}{p}\frac{\partial D}{\partial x}a\cos\left(\frac{x}{p}\right)\right].
 \end{equation}

We can better understand the effects by using a trigonometric identity to express Eqn.~(\ref{eq:ScatEdge}) as a single sinusoid (i.e. $A\sin(x)+B\cos(x)=\sqrt{A^2+B^2}\sin\left(x+\arctan(B/A)\right)$, noting that here we also take a small angle approximation for $\arctan$):

 \begin{equation}
 \label{eq:DiffEdgeSimp}
 I(x,z=\Delta) = a\sqrt{\left(1+\Delta\frac{\partial^2D}{\partial x^2}-\frac{D\Delta}{p^2}\right)^2+\left(\frac{2\Delta}{p}\frac{\partial D}{\partial x}\right)^2}\sin\left(\frac{x}{p}+\frac{2\Delta p \frac{\partial D}{\partial x}}{p^2+\Delta p^2 \frac{\partial^2 D}{\partial x^2}-D\Delta}\right)+b\left(1+\Delta \frac{\partial^2 D}{\partial x^2}\right).
 \end{equation}

This is the diffusion effect we saw in Eqn.~(\ref{eq:SimpleDiffusion}), with some additional effects near an edge, where $\frac{\partial^2D}{\partial x^2}$ is significant, as shown in Fig.~\ref{fig:darkfield} b.  In any edge regions, we see that the amplitude of oscillations in the periodic illumination can now be increased or decreased and that the local mean intensity ($b$) can be increased or decreased, since it is multiplied by $\left(1+\Delta\frac{\partial^2D}{\partial x^2}\right)$.  Because the local first and second spatial derivatives can change quickly (e.g. on the edge of a scattering object), this is a local effect that can result in an illumination that is no longer well-described by a simple sinusoid.

 \begin{figure}
 \centering
 \includegraphics[width=500pt]{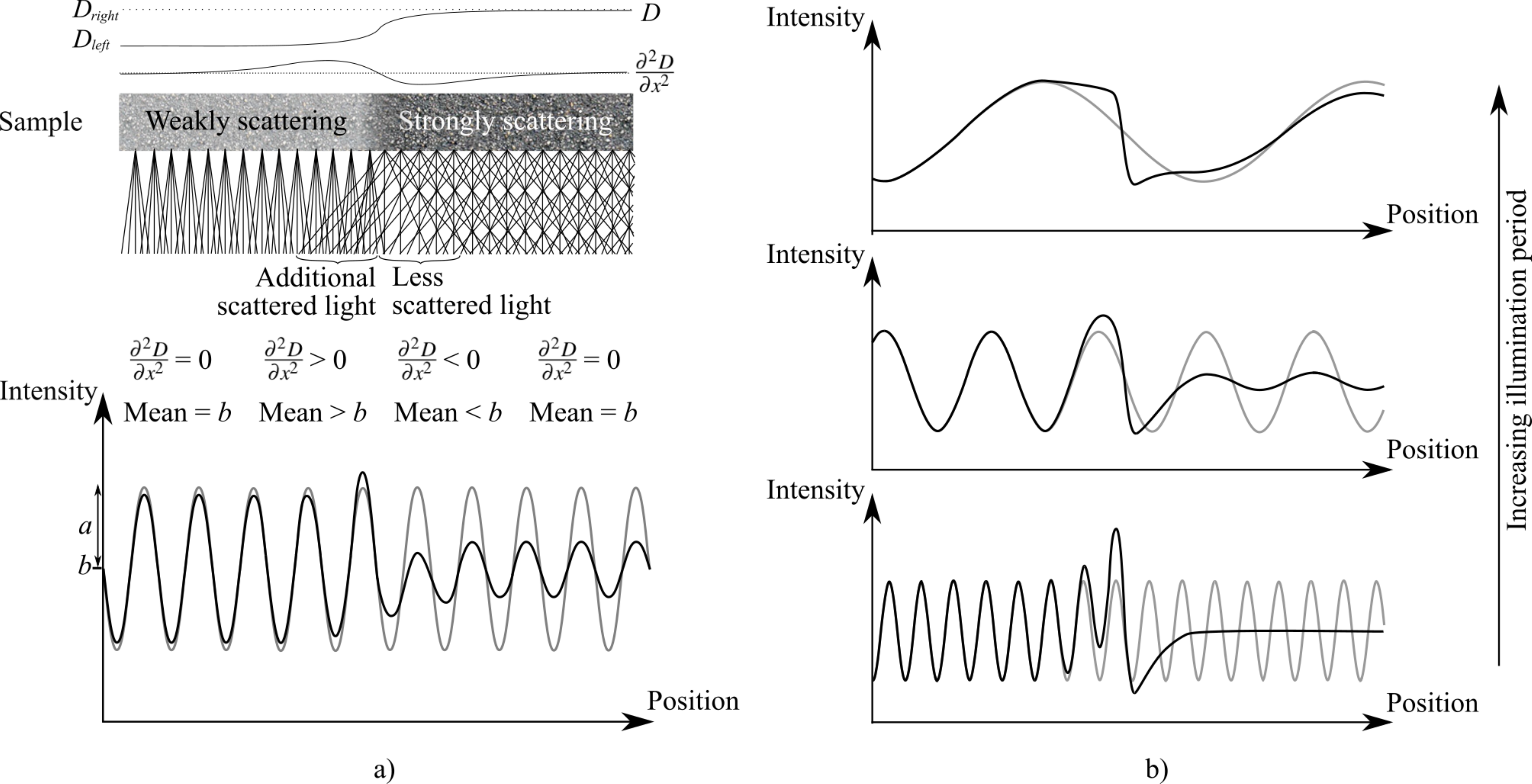}
 \caption{a) We can examine the effects seen where the diffusion or dark-field/scattering property $D$ changes spatially.  The test sample here is weakly scattering on the left and strongly scattering on the right, as modelled by Eqn. (\ref{eq:SmoothEdge}), meaning that $\partial^2D/\partial x^2$ is non-zero during the transition between the two regions. The resulting intensity can be described by Eqn. (\ref{eq:ScatEdge2}) or, if the area on which the scattered light falls ($\sqrt(D\Delta)$) is wider than the illumination period $p$, by Eqn.~(\ref{eq:DiffConvolSol}). The area indicated as `Additional scattered light' in the sample collects light that is scattered from the right, so collects more scattered light than the rest of the area behind the weakly scattering volume.  This means that, relative to the rest of the weakly scattering area on the left, the mean intensity is increased here, and, depending on the period of the illumination relative to the spread of the area, the local visibility of the periodic illumination is increased, since $\partial^2D/\partial x^2$ is positive. The area indicated as `Less scattered light' experiences less scattered light than the rest of the strongly scattering area on the right, since the scatterers in the adjacent weakly scattering area do not scatter as broadly.  This means that, relative to the rest of the area behind the strongly scattering part of the sample, the mean intensity is decreased here, and the visibility of the periodic illumination is decreased, since $\partial^2D/\partial x^2$ is negative.  
 Panel b) shows the effect of changing the illumination period, while the diffusion/dark-field property $D$ is fixed, which helps to reveal the shape of the scattering edge signature. Note also that the reduction in visibility is much more significant when the period of oscillation approaches and becomes smaller than the length scale associated with the diffusion/scattering.}
 \label{fig:diffedgedetail}
 \end{figure}

This can be explored further by looking at two neighbouring regions with differing  scattering properties, setting
\begin{equation}
\label{eq:SmoothEdge}
D(x) = ((D_{right} - D_{left})/2) \textnormal{Erf}[x] + ((D_{right} + D_{left})/2)
\end{equation}

so that $D = D_{right}$ far to the left of the origin, and $D = D_{left}$ to the right of the origin, with some smooth transition between, described by the Erf function (i.e. the special function known as the `error function').

This gives:
\begin{eqnarray}
\label{eq:ScatEdge2}
I(x,z=\Delta) = a \sqrt{\left(1-\frac{\Delta  \left(\frac{1}{2} \text{Erf}(x) (D_{right}-D_{left})+\frac{D_{left}+D_{right}}{2}\right)}{p^2}-\frac{\Delta  \left(2 e^{-x^2} x (D_{right}-D_{left})\right)}{\sqrt{\pi }}\right)^2+\left(\frac{\Delta  \left(e^{-x^2} (D_{right}-D_{left})\right)}{\sqrt{\pi } p}\right)^2} \nonumber
\\ \times
\textrm{ sin} \left(\frac{x}{p}-\frac{\Delta  p \left(e^{-x^2} (D_{right}-D_{left})\right)}{\sqrt{\pi } \left(-\Delta  \left(\frac{1}{2} \text{Erf}(x) (D_{right}-D_{left})+\frac{D_{left}+D_{right}}{2}\right)-\frac{\Delta  p \left(2 e^{-x^2} x (D_{right}-D_{left})\right)}{\sqrt{\pi }}+p^2\right)}\right) \nonumber
\\ + 
b \left(1-\frac{\Delta  \left(2 e^{-x^2} x (D_{right}-D_{left})\right)}{\sqrt{\pi }}\right).
\end{eqnarray}

As mentioned in the final paragraph of section \ref{sec:UniformScat}, the applicability of a solution found using a finite-difference approximation (Eqn.~(\ref{eq:TaylorApprox})) is limited to the case where the area of diffusion is smaller than the period of the illumination pattern. Using the alternative approach given in Eqn. (\ref{eq:PosDepPSF}), we can evaluate first the diffusion on the left of the origin, and then the diffusion effect on the right, with the intensity given by Eqn. (\ref{eq:Illumination}) ($I(x')=a\sin{\left(\frac{x'}{p}\right)+b}$):

\begin{equation}
	I(z=\Delta)=\int_{-\infty }^{0} \left(a \textrm{sin} \left(\frac{x'}{p}\right)+b\right)\frac{e^{-\frac{(x-x')^2}{2 D_{left}\Delta}} }{\sqrt{2 \pi  D_{left} \Delta}} dx' + \int_{0}^{\infty } \left(a \textrm{sin} \left(\frac{x'}{p}\right)+b\right)\frac{e^{-\frac{(x-x')^2}{2 D_{right}\Delta}} }{\sqrt{2 \pi  D_{right} \Delta}} dx'.
    \label{eq:DiffConvol}
\end{equation}

This evaluates to:

\begin{eqnarray}
\label{eq:DiffConvolSol}
I(z=\Delta)=\frac{ia}{4}\mathrm{e}^{-\frac{2ipx + (D_{left}+D_{right})\Delta}{2 p^2}}\left[-\mathrm{e}^{\frac{D_{left}\Delta}{2p^2}}\left(-1-\operatorname{Erf}\left(\frac{px-iD_{right}\Delta}{p\sqrt{2D_{right}\Delta}}\right)+\mathrm{e}^{\frac{2ix}{p}}\left(1+\operatorname{Erf}\left(\frac{px+iD_{right}\Delta}{p\sqrt{2D_{right}\Delta}}\right)\right)\right) \right.\nonumber\\
\left.+\mathrm{e}^{\frac{D_{right}\Delta}{2p^2}}\left(1-\operatorname{Erf}\left(\frac{px-iD_{left}\Delta}{p\sqrt{2D_{left}\Delta}}\right)-\mathrm{e}^{\frac{2ix}{p}}\left(1-\operatorname{Erf}\left(\frac{px+iD_{left}\Delta}{p\sqrt{2D_{left}\Delta}}\right)\right)\right)\right] \nonumber\\
+ 2b\mathrm{e}^{\frac{2ipx+(D_{left}+D_{right})\Delta}{2p^2}}\left(2-\operatorname{Erf}\left(\frac{x}{\sqrt{2D_{left}\Delta}}\right)+\operatorname{Erf}\left(\frac{x}{\sqrt{2D_{right}\Delta}}\right)\right).
\end{eqnarray}

 Equation~\ref{eq:DiffConvolSol} can correctly describe effects seen at an edge without the problems associated with the potential for a `negative' sinusoidal solution given in Eqn.~(\ref{eq:ScatEdge2}). This means that the diffusion effect can be modelled when it extends over several periods of the illumination.  Several cases are shown in Fig.~\ref{fig:diffedgedetail} b, varying the illumination period $p$ only, and these cases help to develop an understanding for the `diffusion edge effect'. The python code in the Supplementary Material implements Eqn.~\ref{eq:DiffConvolSol}, providing a GUI to explore this effect (also at https://github.com/KayeMorgan7/XrayFokkerPlanckSim). These results indicate that the strongest effect is the variations in the mean value (the last term in Eqn.~(\ref{eq:ScatEdge2})), resulting in a signal, shaped like the second derivative of the diffusion coefficient, that sits `on top' of the sinusoidal illumination.  This origin of this intensity variation can be described by considering the areas where `additional scattered light' and `less scattered light' are observed, as shown in Fig.~\ref{fig:diffedgedetail} and described in the caption.

Note, however, that with this solution (Eqn.~(\ref{eq:DiffConvolSol})) the dark-field signal ($D_{left},D_{right}$) cannot be set to zero as this will cause problems associated with dividing by zero, but can be set to a very small number when required.
 
\subsubsection{Modelling scattering using a dark-field coefficient}
It is of interest to have a way to characterise the diffusion characteristic that is consistent with the Fokker--Planck model.  To reach this, we can look at the original differential equation (Eqn.~(\ref{eq:FokkerPlanckXray})) without phase effects, $\phi$=0.  This gives the diffusion equation:

\begin{equation}
\label{eq:FokkerPlanckXrayDiff}
\frac{\partial}{\partial z} I = \frac{\partial^2}{\partial x^2} (D I).
\end{equation}

If we have any periodic illumination (Eqn.~(\ref{eq:Illumination})), and we consider an area over which the diffusion/scattering is not changing significantly with $x$ (so that $D$ can be moved outside of $\frac{\partial^2}{\partial x^2}$), then Eqn.~(\ref{eq:FokkerPlanckXrayDiff}) must satisfy:
\begin{equation}
\label{eq:FokkerPlanckXrayDiff2}
\frac{\partial}{\partial z} I = -\frac{D}{p^2} I.
\end{equation}

\noindent This differential equation can be solved as

\begin{equation}
I(x,z)=I(x,z=0)e^{-\frac{Dz}{p^2}},
\end{equation}

\noindent which agrees with the `linear dark-field co-efficient' of $D/p^2$ seen in Lynch et al. \cite{lynch2011}, and used in dark-field computed tomography \cite{bech2010quantitative}.

\subsection{Phase effects}
\label{sec:Phase}

Looking at a sample with phase properties, $\phi \neq$ 0, independently of dark-field effects (setting $D=0$), from Eqn.~(\ref{eq:FokkerPlanckXray}), we can approximate the intensity observed at distance $z=\Delta$ by the finite-difference approximation (Eqn.~(\ref{eq:TaylorApprox})), insert sinusoidal illumination (Eqn.~(\ref{eq:Illumination})) and evaluate the derivative with respect to $x$ to give

\begin{equation}
\label{eq:PhaseOnly}
I(x,z=\Delta) =a\sin\left(\frac{x}{p}\right) + b - \frac{\Delta}{k}\left[\left(a\sin\left(\frac{x}{p}\right)+b\right)\frac{\partial^2 \phi}{\partial x^2}+\frac{1}{p}a\cos\left(\frac{x}{p}\right) \frac{\partial \phi}{\partial x}\right].
\end{equation}

Using the trigonometric identity for the addition of two sinusoidal signals of the same argument (provided immediately before Eqn.~(\ref{eq:DiffEdgeSimp})), and leaving out terms of higher order $\Delta$ (since we linearised with $\Delta$ in Eqn.~(\ref{eq:TaylorApprox})), Eqn.~(\ref{eq:PhaseOnly}) simplifies to:

\begin{equation}
\label{eq:PhaseSimp}
I(x,z=\Delta) =\left(1-\frac{\Delta}{k}\frac{\partial^2\phi}{\partial x^2}\right)\left[a\sin\left(\frac{x}{p}-\tan^{-1}\left(\frac{\frac{\partial \phi}{\partial x}}{p\left(\frac{k}{\Delta}-\frac{\partial^2 \phi}{\partial x^2}\right)}\right)\right)+b\right].
\end{equation}

\subsubsection{Prism effect, seen for uniform phase gradient \texorpdfstring{$\frac{\partial \phi}{\partial x}$}.}
\label{sec:prism}
If we are in the slowly-changing bulk of the sample, where $\frac{\partial\phi}{\partial x}$ is significant, but $\frac{\partial^2\phi}{\partial x^2}$ is small and can be neglected, Eqn.~(\ref{eq:PhaseSimp}) simplifies to:

\begin{equation}
\label{eq:PhasePrism}
I(x,z=\Delta) = a\sin\left(\frac{x}{p}-\tan^{-1}\left(\frac{\Delta} {p k}\frac{\partial\phi}{\partial x}\right)\right) + b.
\end{equation}

This means that if the phase shift is linear (so $\frac{\partial^2\phi}{\partial x^2}$ is zero), then we see only the prism effect, where the grid pattern shifts transversely.  The shift in the relative position is what we would expect (see Morgan et al.\cite{morgan2011quantitative}), namely $\frac{\Delta}{k}\frac{\partial\phi}{\partial x}$), and there is no change in the mean or visibility of the signal. This is the `well-behaved' case, where phase effects are not introducing any signals that look like attenuation or dark-field.

\begin{figure}
\centering
\includegraphics[width=500pt]{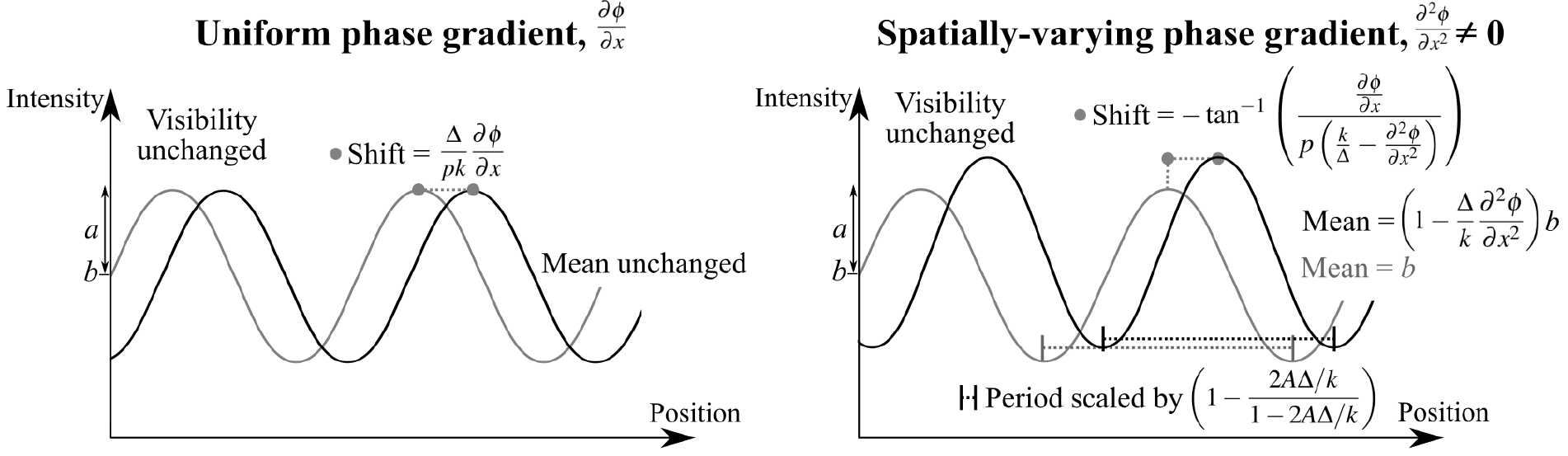}
\caption{a) A shift of the illumination is the basis for a phase shift signal, shown here for the simple case where the second derivative of the phase gradient is not significant. The shift in the transverse position of the periodic illumination is described by Eqn.~(\ref{eq:PhasePrism}). b) When the second derivative of the phase gradient is significant, and the phase can be locally described as $\phi=Ax^2+Bx+c$, we can see a transverse shift, squeezing and a local increase/decrease of the periodic illumination, described by Eqn.~(\ref{eq:PhaseSimp})/(\ref{eq:QuadPhase}).}
\label{fig:PhaseGradient}
\end{figure}

\subsubsection{Focusing effect, seen for \texorpdfstring{$\frac{\partial^2 \phi}{\partial x^2} \neq 0$})}
\label{sec:focusing}
If the phase gradient is not uniform, and varies with position $x$, then we will have more interesting effects.  We can consider each of the terms in Eqn.~(\ref{eq:PhaseSimp}) to understand the physical meaning.  With soft x-rays and a propagation distance in the near field, $\frac{\Delta}{k}$ will be very small (e.g. $10^{-11}$). If the second phase derivative is large, i.e. approaching the order of $\frac{k}{\Delta}$, then the multiplicative term at the front of the sinusoid in Eqn.~(\ref{eq:PhaseSimp}) can be slightly bigger or smaller than 1, describing either some amplification or deamplification of the entire illumination.  Whether amplification or deamplification is seen will depend on the sign of the second derivative of the phase, which describes either defocusing or focusing of the illumination.  We also see that if the second phase derivative approaches the order of $\frac{k}{\Delta}$, then the denominator inside the $\tan^{-1}$ in Eqn.~(\ref{eq:PhaseSimp}) will become slightly bigger or smaller than $\frac{k}{\Delta}$, describing either a squeezing or a spreading out of the intensity peaks, matching with the defocusing/focusing behaviour.

To untangle the effect of phase gradients on the period of the illumination, from the effects of phase gradients on the shift of the illumination (i.e. the $x$-dependent terms within the sinusoid), we can model the case of some `phase curvature' using a Taylor polynomial of order two, $\phi=Ax^2+Bx+C$, so that $\frac{\partial\phi}{\partial x}=2Ax+B$ and $\frac{\partial^2 \phi}{\partial x^2}=2A$.  The values for $A$, $B$ and $C$ could be chosen to describe the phase gradient and curvature of any local area of the sample.  With this phase, Eqn.~(\ref{eq:PhaseSimp}) becomes:

\begin{eqnarray}
\label{eq:QuadPhase}
I(x,z=\Delta) &=&\left(1-\frac{2A\Delta}{k}\right)\left[a\sin\left(\frac{x}{p}-\tan^{-1}\left(\frac{(2Ax+B)}{p\left(\frac{k}{\Delta}-2A\right)}\right)\right)+b\right]
\nonumber
\\
&\approx&\left(1-\frac{2A\Delta}{k}\right)\left[a\sin\left(\frac{x}{p}\left(1-\frac{2A\Delta/k}{1-2A\Delta/k}\right)+\frac{B\Delta}{pk(1-2A\Delta/k)}\right)+b\right],
\end{eqnarray}

\noindent with the simplification taken for small shifts in phase of the sinusoid, so that $\tan(x)\approx x$.

\begin{figure}
\centering
\includegraphics[width=380pt]{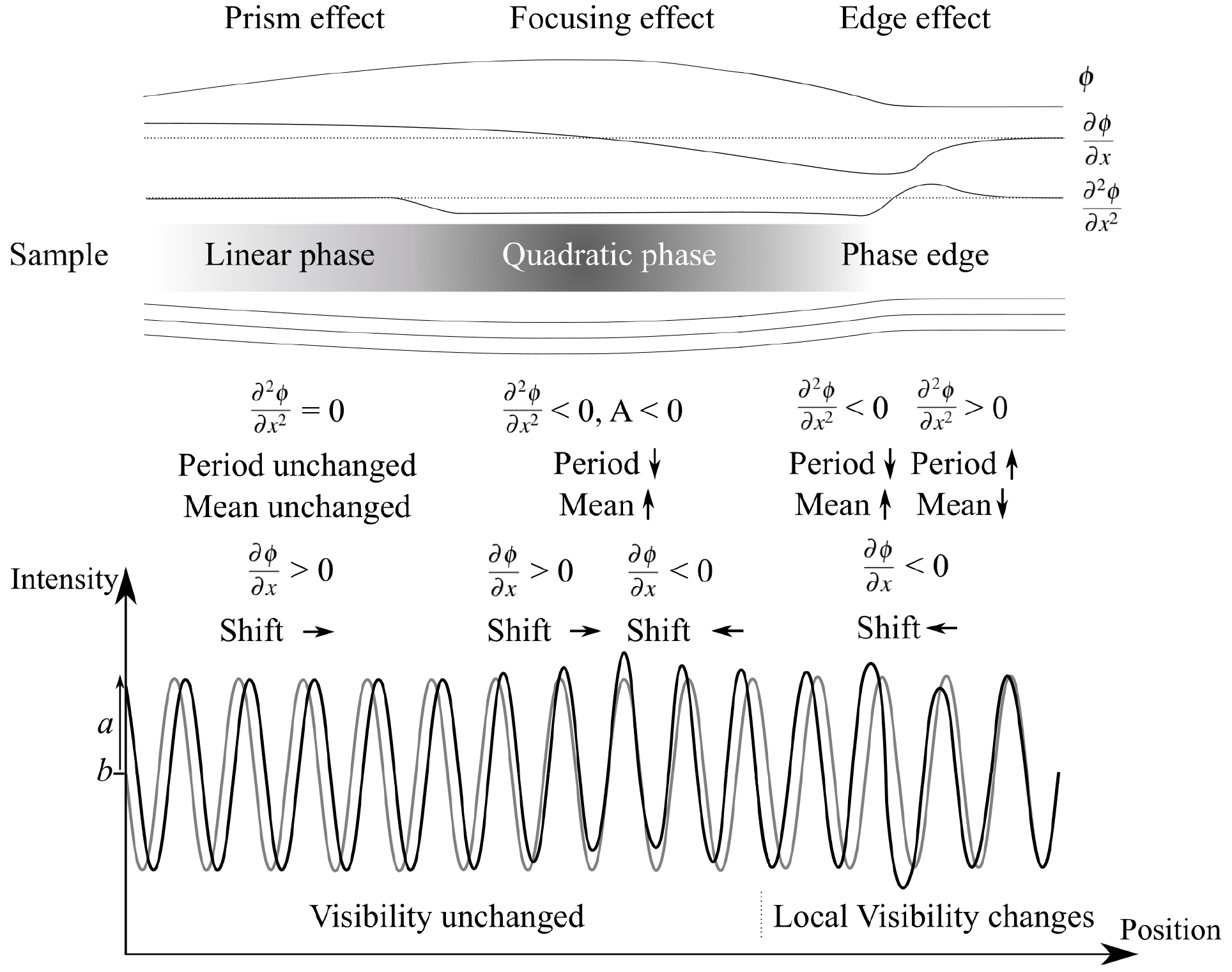}
\caption{Phase variations incurred by the sample result in changes to the sinusoidal illumination in a number of ways, shown here (left to right) for a linear projected phase (Fig.~\ref{fig:PhaseGradient} a), Section \ref{sec:prism}), which shifts the illumination transversely, a quadratic projected phase (Fig.~\ref{fig:PhaseGradient} b), Section~\ref{sec:focusing}), which focuses the illumination, locally increasing the mean intensity and a phase edge (Section~\ref{sec:phaseedge}), which introduces an additional bright/dark variation to the illumination.}
\label{fig:phasecurvature}
\end{figure}

We see that a linear change in phase (modelled by $A=0$, $B\neq0$) will shift the relative position of the peaks of the periodic illumination by $\frac{B\Delta}{k}$, acting as a prism without changing the period of the illumination, as seen in Eqn.~(\ref{eq:PhasePrism}), and on the left third of Fig.~(\ref{fig:phasecurvature}).

If the phase shift incurred by the sample is not just linear, but there is such curvature that a quadratic term is required to adequately describe the local phase, then we will see additional effects.  This case corresponds to when $A$ is significant, approaching $\frac{k}{2\Delta}$, so that the term $1-\frac{2A\Delta}{k}$ moves away from 1.  Since this term appears in three places, a large $A$ will 1) change the mean illumination, 2) alter the period and 3) introduce an additional shift in the position of the periodic illumination.  If the phase curvature is positive (seen for $A>0$ in our quadratic model) then the beamlets will spread out, seen by a decreased mean illumination and a period that is multiplied by a number slightly greater than 1 and so locally increased.  The shift in the position of the illumination will also be slightly reduced towards the centre of the focus.  If the phase curvature in negative ($A<0$), the illumination will be focused, increasing the mean intensity and decreasing the period of the periodic illumination, as shown in Fig.~\ref{fig:phasecurvature}.  These observations are consistent with preservation of photons.  While this kind of focusing will not introduce any visibility changes when the grid pattern is directly resolved, it can change the visibility measured in grating interferometry \cite[Fig.1]{koenig2016origin} (See section \ref{sec:Interferometry}).

\subsubsection{Phase edge, seen for \texorpdfstring{$\frac{\partial^2 \phi}{\partial x^2} \neq 0$})}
\label{sec:phaseedge}
A phase edge is also an interesting feature to study, particularly given the bright/dark contrast seen in propagation-based phase contrast imaging when a wave passes across an edge and self-interferes.  If we model a phase edge in the same way as Eqn.~(\ref{eq:SmoothEdge}) models a scattering edge (replacing $D$ with $\phi$) and use this model in our expression for phase effects, Eqn.~(\ref{eq:PhaseSimp}), we get

\begin{eqnarray}
\label{eq:PhaseEdge}
I(x,z=\Delta) =\left(\frac{\Delta  \left(2 e^{-x^2} x (\phi_{right}-\phi_{left})\right)}{\sqrt{\pi } k}+1\right) \left(a \sin \left(\left(\frac{x}{p}-\tan ^{-1}\left(\frac{e^{-x^2} (\phi_{right}-\phi_{left})}{p \sqrt{\pi } \left(\frac{k}{\Delta }+\frac{2 e^{-x^2} x (\phi_{right}-\phi_{left})}{\sqrt{\pi }}\right)}\right)\right)\right)+b\right).
\end{eqnarray}

When plotted, as expected, a bright/dark contrast fringe overlies the illumination, in a similar manner to the dark-field edge. The python code in the Supplementary Material implements Eqn.~\ref{eq:PhaseEdge}, providing a GUI to explore this effect (also at https://github.com/KayeMorgan7/XrayFokkerPlanckSim). This can result in local changes to the visibility, as seen in Fig.~\ref{fig:phasecurvature}. 

\subsection{Combined effects}

While the phase effects and dark-field effects are most easily understood in the previous two sections, where each is isolated, we can also write an equation to describe the combined effects.  Since the shifting and diffusive terms in the finite-difference pproximation for the Fokker--Planck equation (Eqn.~(\ref{eq:TaylorApprox})) add together, we can describe the resulting intensity as the sum of Eqn.~(\ref{eq:DiffEdgeSimp}) and Eqn.~(\ref{eq:PhaseSimp}), subtracting the illuminating intensity (Eqn.~(\ref{eq:Illumination})) since it is in both equations and hence repeated.  This gives:

 \begin{eqnarray}
 \label{eq:AllEffects}
 I(x,z=\Delta) &=& a\sqrt{\left(1+\Delta\frac{\partial^2D}{\partial x^2}-\frac{D\Delta}{p^2}\right)^2+\left(\frac{\Delta}{p}\frac{\partial D}{\partial x}\right)^2}\sin\left(\frac{x}{p}-\frac{\Delta p \frac{\partial D}{\partial x}}{p^2+\Delta p \frac{\partial^2 D}{\partial x^2}-D\Delta}\right)+b\left(1+\Delta \frac{\partial^2 D}{\partial x^2}\right) 
 \nonumber 
 \\ &&+
 \left(1-\frac{\Delta}{k}\frac{\partial^2\phi}{\partial x^2}\right)\left[a\sin\left(\frac{x}{p}-\tan^{-1}\left(\frac{\frac{\partial \phi}{\partial x}}{p \left(\frac{k}{\Delta}-\frac{\partial^2 \phi}{\partial x^2}\right)}\right)\right)+b\right] 
 \nonumber 
 \\ &&-
 \left(a\sin\left(\frac{x}{p}\right)+b\right).
 \end{eqnarray}

\subsection{Extension of the Fokker--Planck model to Grating Interferometry and Edge Illumination}
\label{sec:InterferometryEdgeIllumination}

The results presented in the preceding sections can be extended to include the effect of an analyser grating or detector mask, as would be used in a grating interferometer or edge illumination set-up respectively. The use of a second grating/mask has the advantage that larger pixels can be used, which is typically more efficient and allows the imaging of larger samples than imaging where the illumination pattern is directly resolved. To incorporate the second grating/mask, we can simulate a periodic absorption immediately before the detector to `analyse' the illumination we have studied so far.  

Here we show the implementation of this, using a general signal that describes any of the cases we have explored (Eqn.~(\ref{eq:ScatEdge2}), Eqn.~(\ref{eq:PhasePrism}), Eqn.~(\ref{eq:QuadPhase}), Eqn.~(\ref{eq:PhaseEdge}) etc.), where the sinusoidal illumination arriving at the periodic absorber has been modified by a factor of $M$ in mean value, a factor of $V$ in visibility, and has been shifted by $s$, so can be described as

\begin{equation}
\label{eq:ModifiedSinusoid}
I(x,z=\Delta)=MV \sin\left(\frac{2 \pi (x-s)}{p}\right)+M. 
\end{equation}

Note that we neglect a local change in period here, due to the increased complexity, but the same approach can be applied to study this effect.  Also note that we have explicitly included a factor of $2\pi$ inside the sinusoid in Eqn. \ref{eq:AbsorptionGrating} that is not included earlier, so that it is possible to integrate the unresolved intensity oscillations over the period of that intensity oscillation, which will now correspond exactly to $p$.  We can model the absorption by the analyser grating or detector mask by multiplying the illumination (Eqn.~(\ref{eq:ModifiedSinusoid})) by a sinusoid that spatially oscillates between 0 (total absorption) and 1 (total transmission) (i.e. $\frac{1}{2} \sin \left(\frac{2 \pi  (n-x)}{p}\right)+\frac{1}{2}$). 

\subsubsection{Grating Interferometry}
\label{sec:Interferometry}

In grating interferometry, the intensity is recorded during a `stepping scan', where an absorption grating steps transversely across the illumination.  This can be implemented by convolving the `absorption' sinusoid with the modified illumination sinusoid (Eqn.~(\ref{eq:ModifiedSinusoid})) over the period of the illumination, $p$, providing the infinitely sampled measured stepping curve as a function of absorption grating position $n$ (which should be sampled from $n=0$ to $n=p$),

\begin{eqnarray}
\label{eq:AbsorptionGrating}
\textrm{Stepping curve}(n) &=& \int_0^p \left(\frac{1}{2} \sin \left(\frac{2 \pi  (n-x)}{p}\right)+\frac{1}{2}\right) \left(M V \sin \left(\frac{2 \pi  (x-s)}{p}\right)+M\right) dx
\\
&=& \frac{p}{4} MV \cos \left(\frac{2 \pi  (n-s)}{p}\right)+\frac{p}{2} M.
\end{eqnarray}

This stepping curve has a mean value of $Mp/2$, a visibility of $V/2$ and is shifted by $s$.  This means that the results shown above for a directly-resolved grid can also be applied in the case of an interferometer with just a simple multiplicative scaling of $p/2$ for the mean (attenuation), $1/2$ for the visibility (dark-field) and no adjustment for the shift. 

Note that if the period of oscillations is locally varying (seen when $\frac{\partial^2 \phi}{\partial x^2} \neq 0$, Sections~\ref{sec:focusing} \& \ref{sec:phaseedge}), we would need to scale the period $p$ by a factor of $f$ in the general expression for a modified sinusoid, Eqn.~(\ref{eq:ModifiedSinusoid}). Carrying this into Eqn.~(\ref{eq:AbsorptionGrating}) produces a non-trivial solution that depends on the nature of $\frac{\partial^2 \phi}{\partial x^2}$, in a similar way to the results seen in Sections~\ref{sec:focusing} and \ref{sec:phaseedge}, where we had to consider specific cases of $\frac{\partial^2 \phi}{\partial x^2}$. Including spatial variations in illumination period would be important when considering how a strong phase object measured using interferometry may produce a `pseudo' dark-field (e.g. \cite[Fig.~1]{koenig2016origin}).

\subsubsection{Edge Illumination}

Edge illumination imaging has several variants, for example, depending on whether there is strong scattering expected from the sample\cite{olivo2007coded, Endrizzi2018}. With the aim only to illustrate the applicability of the Fokker--Planck approach, here we will consider just the simple case where scattering is weak. In this case, the absorption grating / detector mask, which has a period matching the illumination period, should be aligned so that the peak of the sinusoidal illumination first coincides with the left edge of the each absorption line (exposure 1, $I_{left}$), then coincides with the right edge of the absorption line (exposure 2, $I_{right}$),

\begin{eqnarray}
I_{left} &=& \int_0^p \left(\frac{1}{2} \cos \left(\frac{2 \pi  x}{p}\right)+\frac{1}{2}\right) \left(M V \sin \left(\frac{2 \pi  (x-s)}{p}\right)+M\right) dx
\nonumber
\\
\label{eq:EdgeIllum1}
&=&-\frac{1}{4} M p V \sin \left(\frac{2 \pi  s}{p}\right)+\frac{M p}{2}
\\
I_{right} &=& \int_0^p \left(-\frac{1}{2} \cos \left(\frac{2 \pi  x}{p}\right)+\frac{1}{2}\right) \left(M V \sin \left(\frac{2 \pi  (x-s)}{p}\right)+M\right) dx
\nonumber
\\
\label{eq:EdgeIllum2}
&=& \frac{1}{4} M p V \sin \left(\frac{2 \pi  s}{p}\right)+\frac{M p}{2}
\end{eqnarray}

The shift ($s$) and attenuation ($M$) signals could then be extracted by adding and subtracting Eqns.~(\ref{eq:EdgeIllum1}) and (\ref{eq:EdgeIllum2}), giving consistent results as in the scenario described above for grating interferometry.

Following these demonstrations of applicability, we leave further applications of the Fokker--Planck equation to Grating Interferometry and Edge Illumination to future work.  Possibilities include integrating over a non-integer number of grating lines and investigating the effects of a locally-modified period.

\section{Discussion}

This work serves as a starting point for future applications of the Fokker--Planck and Kramers--Moyal equations in x-ray imaging techniques that capture phase and dark-field effects. We begin the discussion by mentioning a number of possibilities for future work.  A clear next step in any grating-based or speckle-based approach is to utilise this model as a starting point for phase/dark-field retrieval, or simply as a basis for lensing and edge-based `corrections' to retrieved attenuation, phase and dark-field signals (perhaps as a second or iterative step after the initial extraction of the three signals by conventional retrieval methods).  As noted in Section~\ref{sec:Interferometry}, the Fokker--Planck model could be applied to provide an analytical model for specific cases seen in grating interferometry where phase edge\cite{yashiro2015} and lensing effects (simulated and measured in Wolf et al.\cite{wolf2015lens}) result in an illumination period that is locally-varying, leading to a spatially-varying `mis-match' between that illumination and the second/analyser grating\cite{koenig2016origin}.  

Corrections for dark-field-edge, phase-edge and lensing effects are particularly of interest when the size of the features within the sample are comparable to the period of the illumination, and the effects are typically more dominant.  The Fokker--Planck model described in this paper could better shed light on the transition between two regimes, the regime where sample features are large enough to create a spatially-resolved phase signal (e.g.~see differential phase images of individual air sacs in the lung in Morgan et al.~\cite{morgan2016} or high-resolution lung CT in Koenig et al.~\cite[Fig.5]{koenig2016origin}) and the regime where sample features are not spatially resolved and are instead seen as a dark-field signal (e.g. see dark-field images of the lungs\cite{schleede2012emphysema} or \cite[Fig.2]{koenig2016origin}). The Fokker--Planck model naturally lends itself to a parameter-based definition for these regimes, differentiating between edge effects and bulk effects, as determined by the spatial resolution of the imaging set-up and the illumination period.
This direction of future work also lends itself to an investigation into the optimum illumination pattern period to be sensitive to a given dark-field signal for a given wavelength with a given imaging set-up. This links to the correlation length, $\lambda \Delta / p$ seen in grating interferometry \cite{prade2016}, which matches the expression seen when the `smear width' $L$ is set to the illumination period $p$, zeroing the visibility of the resulting illumination at the detector (also see Fig.~3 c) of the companion paper on the X-ray Fokker--Planck equation \cite{Paganin2019}).

There are also natural extensions to the Fokker--Planck model. The first extension would be to describe more detailed\cite{modregger2012} and size-related\cite{lynch2011, modregger2017} scattering functions. A dark-field signal from a non-uniform collection of overlaid and unresolved sample features could be modelled by the addition of extra diffusive terms ($\partial^2(D_mI)/\partial x^2$, each with a different $D_m$) to the Fokker--Planck equation (Eqn.~(\ref{eq:FokkerPlanckXray})). Since the diffusion equation is energy-conserving, the addition of these terms would require no other adjustments.  A dark-field signal from a collection of extended unresolved sample features (e.g. carbon fibres \cite{revol2013, kagias2016} or bone microstructure \cite{jensen2010directional, jud2017}) could be modelled by a two-dimensional Fokker--Planck equation (a truncation of the Kramers--Moyal equation), which provides three parameters to describe the two-dimensional scattering ellipse utilised in directional dark-field imaging \cite{jensen2010prb} (which is sometimes modelled as scattering that oscillates in magnitude sinusoidally with angle \cite{jensen2010directional, potdevin2012x}). 

The Fokker--Planck model shown here could be extended to incorporate real-world effects seen at synchrotron and laboratory x-ray sources, including limited spatial resolution, extended sources, polychromatic illumination and magnification. The spatial resolution limit due to the detector characteristics, including pixel size, could be modelled by convolving the general expressions derived above (e.g. Eqn.~(\ref{eq:AllEffects})) with a Gaussian of specified width. The spatial resolution associated with an x-ray imaging system will also be limited by the size of the x-ray source, with an extended source `blurring' the observed image.  In the Fokker--Planck model, an extended source could be modelled by summing the images created by an array of many point sources, produced by shifting the intensity measured at $z=\Delta$ slightly in order to model the grid projection seen from each point on the source.  An alternative would be to replace $D(x,y)$ with $D(x,y)+d$ (or equivalently, add another diffusion term), where $d$ models the extra blur that results from source-size blurring (see p. 13 of Paganin \& Morgan~\cite{Paganin2019}). The effect of an extended source is also of particular relevance in the measured x-ray dark-field signal \cite[Sec.5]{koenig2016origin}. In the direction of less blurring, a more finely-resolved structured illumination behind a rectangular (or some other-shaped \cite{yaroshenko2014}) grid could be modelled using a Fourier sum of the sinusoidal illumination model utilised in this paper.

Any x-ray source that is not a synchrotron will typically produce illumination that is sufficiently polychromatic and divergent that the associated effects must be included in modelling. A divergent source could be modelled according to the Fresnel scaling theorem \cite{Paganin2006}, by setting the propagation distance $z$ to $z/M$, where $M$ is the magnification, while adjusting the native pixel size accordingly.  This is also key in x-ray dark-field imaging by interferometry \cite{yashiro2018effect} \cite[Sec.4]{koenig2016origin}).  An image captured with polychromatic illumination could be modelled as the weighted sum of the images seen at each wavelength within the illuminating spectrum \cite{wilkins1996phase}.  This can be implemented by integrating the Fokker--Planck equation (Eqn.~(\ref{eq:TaylorApprox})) over wavelength, after multiplying Eqn.~(\ref{eq:TaylorApprox}) with a wavelength-dependent weighting factor that describes the x-ray spectrum and which could also incorporate detector sensitivity. 
 
A sum of many slightly-different images into a single exposure can be used to describe a reduction in fringe or patterned illumination visibility for a number of cases. In addition to the cases of an extended source and a polychromatic source, reduced visibility is seen when using a diffuser.  A diffuser is typically a sheet of highly-scattering material, like sandpaper, which is placed in the x-ray beam upstream of the sample and rotated at high speed, such that many positions are summed over the time of a typical x-ray exposure.  This will reduce the visibility of propagation-based x-ray phase contrast fringes, as each diffuser position slightly distorts the fringes, and the sum over all diffuser positions results in a fringe that is `blurred' \cite{irvine2010}.  A diffuser will not change the spatially coherent width/area, but will reduce the magnitude of the second order degree of coherence \cite{morgan2010measurement}. In x-ray SAXS, the wavefield is locally distorted by scattering from sub-pixel structures in the sample, locally reducing the spatially coherent width/area, and resulting in a reduction in the visibility of fine features like the illuminating grid pattern. In the case of a diffuser, there is a temporal sum of many slightly distorted images.  In the case of x-ray SAXS, there is a temporal sum of many photons, where some are slightly scattered.

Another aspect of dark-field is worth discussing.  Several papers have reported on an enhancement of dark-field in the vicinity of the boundary of the geometric shadow of a paraxially-illuminated compact object \cite{morrison1992, suzuki1995, yashiro2015}.  Rather than being dark-field due solely to unresolved micro-structure, as is the case for points that lie inside the geometric shadow, the dark-field signal emanating from the edge of the object is augmented by the Young--Maggi--Rubinowicz boundary diffraction wave\cite{BornWolf}.  This phenomenon, whereby illuminated objects have their projected edges acting as a secondary source of (scattered) boundary waves, was first noted in a qualitative study by Young in 1802\cite{YoungOnTheBoundaryWave}.  Subsequent contributions to the underpinning wave theory are due to Maggi\cite{Maggi}, Rubinowicz\cite{Rubinowicz} and Miyamoto \& Wolf\cite{MiyamotoWolf1,MiyamotoWolf2}.  The same phenomenon may also be understood from the perspective of complex-ray theory.  This ray-based perspective underpins Keller's geometric theory of diffraction\cite{Keller}, in which Fresnel diffraction patterns may be understood as the interference between (i) the complex-ray disturbance transmitted by an object according to geometrical optics, and (ii) the complex-ray disturbance scattered from the edges of the object.  A third means of understanding the dark-field-enhancement along the boundary of the geometric shadow of an illuminated object, is from the asymptotic expansion (short-wavelength limit) of the two-dimensional diffraction integral associated with a paraxially-illuminated compact object under the projection approximation\cite{Paganin2006}, obtained by applying the method of stationary phase for double integrals.  Here, the boundary wave corresponds to critical points of the second kind: see e.g.~Sec.~3.3.3 of the text by Mandel \& Wolf\cite{MandelWolf}.  For applications of the boundary-wave concept to coherent X-ray optics, see e.g.~Morgan et al.\cite{Morgan2010a,Morgan2010b}.

\section{Conclusion}

This work has shown that the Fokker--Planck equation can be applied to phase contrast and dark-field x-ray imaging captured using a single-grid set-up, a grating interferometer, an edge-illumination set-up or a speckle-based set-up. The Fokker--Planck model was applied to provide an analytic expression for the additional signals seen as a result of edges within the sample and lensing by the sample.  The authors hope that this work will provide a basis for further applications of the Fokker--Planck model in x-ray phase and dark-field imaging.

\bibliography{sample}

\section*{Acknowledgements}  We acknowledge useful discussions with Florian Schaff, Mario Beltran, Carsten Detlefs, Timur Gureyev, Alexander Kozlov, Thomas Leatham and Tim Petersen. We acknowledge the following funding sources: ARC Future Fellowship FT180100374; Veski Victorian Postdoctoral Research Fellowship (VPRF); German Excellence Initiative and European Union Seventh Framework Program (291763). The python GUI is built on the slider demo from matplotlib, available at \textit{https://matplotlib.org/3.1.1/gallery/widgets/slider\_demo.html}.

\section*{Author contributions statement}

K.S.M. and D.M.P. worked on this paper in close collaboration. K.S.M. prepared all figures and performed all mathematical calculations, arising from discussions between both authors.  The paper was mainly written by K.S.M., with input from  D.M.P.

\section*{Additional information}

{\bf Competing interests:} The authors declare no competing  interests.

%To include, in this order: \textbf{Accession codes} (where applicable); \textbf{Competing interests} (mandatory statement). 

%The corresponding author is responsible for submitting a \href{http://www.nature.com/srep/policies/index.html#competing}{competing interests statement} on behalf of all authors of the paper. This statement must be included in the submitted article file.

\end{document}